\documentclass[pra,twocolumn,showpacs,superscriptaddress]{revtex4-1}
\usepackage{amsmath,amscd,amssymb,color,amsthm}
\usepackage{graphicx,amsfonts,dsfont}
\usepackage{epstopdf}
\usepackage{hyperref}
\usepackage{enumerate,bbold}
\usepackage{array, comment, bm, braket}

\usepackage{color}
\begin{document}
\title{Ground and excited 
state energy calculations of the H$_\mathbf{2}$ molecule 
using a variational quantum eigensolver 
algorithm on an NMR quantum simulator}
\author{Dileep Singh}
\email{dsingh.iisermohali@gmail.com}
\affiliation{Department of Physical Sciences, Indian
Institute of Science Education \& 
Research Mohali, Sector 81 SAS Nagar, 
Manauli PO 140306 Punjab, India.}
\author{Shashank Mehendale}
\email{shashank.mehendale@mail.utoronto.ta}
\affiliation{Department of Chemistry, 
University of Toronto, Toronto, ON M5S 1A1, Canada}
\author{Arvind}
\email{arvind@iisermohali.ac.in}
\affiliation{Department of Physical Sciences, Indian
Institute of Science Education \& 
Research Mohali, Sector 81 SAS Nagar, 
Manauli PO 140306 Punjab, India.}
\author{Kavita Dorai}
\email{kavita@iisermohali.ac.in}
\affiliation{Department of Physical Sciences, Indian
Institute of Science Education \& 
Research Mohali, Sector 81 SAS Nagar, 
Manauli PO 140306 Punjab, India.}
\begin{abstract}
Variational quantum algorithms are emerging as promising candidates for
near-term practical applications of quantum information processors, in
the field of quantum chemistry.  We implement the variational quantum
eigensolver algorithm to calculate the molecular ground-state energy of
the H$_2$ molecule and experimentally demonstrated it on an NMR quantum
processor. Further, we simulate the excited states of the H$_2$
molecule using the variational quantum deflation algorithm and
experimentally demonstrate it on the same NMR quantum processor.  We
also develop the first simulation of the energy calculation of the
H$_2$ molecule using only a single qubit, and verify the results on an NMR
quantum computer. Our experimental results demonstrate that
only a single NMR qubit suffices to calculate the molecular
energies of the H$_2$ molecule to the desired accuracy.
\end{abstract} 
\maketitle 
\section{Introduction}
\label{sec:intro}
The current century is witnessing what is termed as the ``second quantum
revolution'' and quantum computing is at its forefront~\cite{farlane-ptrs-03}.
In recent years, it has been shown that there are a number of problems that a
quantum algorithm would solve much more efficiently as compared to its classical
counterpart~\cite{nielsen-book-10}.  One such class of problems is `quantum
simulation', wherein the goal is to simulate a quantum system efficiently and
gain the maximum amount of information about the
system~\cite{feynman1-ijtp-82}. 

A major class of problems coming under the umbrella of quantum
simulation\cite{lloyd-sci-96} is the problem of finding the eigenstates of the
Hamiltonian of a given system, which has been a prime focus in the field of
quantum chemistry~\cite{Cao-cr-19}.  Finding the ground and excited state
energies of a molecule gives enormous information about its properties, such as
its stability, rate of reactions and the molecular orbitals
involved~\cite{guzik-s-05}.  The calculation of ground and excited state
energies of molecules  is of great importance in chemistry, but this becomes a
challenging task for a classical computer as the complexity of the molecules
grows.  These problems are extremely difficult to solve analytically and hence
the only way out is a numerical approach. However, even this becomes a daunting
task for molecules with a large number of atoms and electrons, due to the large
number of degrees of freedom involved. This is where quantum algorithms
demonstrate their supremacy over their classical counterparts, and various
quantum algorithms have been developed to efficiently calculate the energies of
molecules on a quantum computer~\cite{whitfield-mp-11, kassal-arpc-11,
jones-njp-12}.

Quantum algorithms are prone to a lot of errors and require quantum error
correction which limits their usefulness.  Taking into consideration that we
are in the NISQ era \cite{preskill-quantum-18}, a new class of algorithms has
been developed, which are partly classical and partly quantum, which can reduce
the required gate depth and mitigate errors to a certain extent.  
Variational
Quantum Eigensolver (VQE) is one such algorithm that has been developed to
calculate the ground state of Hamiltonians on a quantum computer
\cite{jarrod-njp-2016,moll-qst-2018,jones-pra-19,wang-prl-19}.  The VQE
algorithm may enable near-term quantum-enhanced
computation~\cite{peruzzo-nature-14} due to its low circuit depth.
 
In addition to ground state energy calculation of molecules, the excited state
energies have several important applications. There have been a few proposals
to develop an algorithm for the detection of excited states of a molecular
Hamiltonian: a method that minimizes the
von Neumann
entropy~\cite{santagati-sa-18} and the quantum subspace expansion
method~\cite{colless-prx-18}. Recently, an algorithm which is an extension of
the VQE algorithm was proposed, namely the Variational Quantum Deflation (VQD)
algorithm, to calculate the excited state energies of
molecules~\cite{higgott-quan-19}.  
The VQD algorithm systematically finds excited states at
almost no extra cost as compared to the other two methods. Using the VQD
algorithm, the excited states are obtained by applying the VQE algorithm to a
modified Hamiltonian which has the excited states as its ground state. Several
experiments has been performed to demonstrate the energy spectra of molecules
using VQE
algorithms
on different quantum platforms
such as photonic quantum processors, trapped ions and
using superconducting qubits~\cite{malley-prx-16,hempel-prx-18,motta-cs-23,jones-pra-24,misiewicz-jpca-24}.

In this paper, we have simulated the ground state energy and excited state
energies of the H$_2$ molecule using the VQE and VQD algorithms, respectively.
The simulated results are then verified on an NMR quantum processor.  To use
the variational principle, we varied the states in the qubit space so as to
find the state which minimizes the energy expectation value. This is done by
using the technique of Unitary Coupled Cluster Singles and Doubles
(UCCSD)~\cite{Xia-iop-20}. The experimental verification is done by calculating
the expectation values terms involved in the H$_2$ molecule Hamiltonian by
using two NMR qubits. The experimental verification involves the calculation of
expectation values of the state, and this is achieved by measuring the
single-qubit Pauli $Z$ operator of the experimentally prepared states. We also
simulated the energies of the H$_2$ molecule on a single qubit system and
verified the results on an NMR quantum computer.  This is the first
experimental demonstration of energy calculations of the H$_2$ molecule which
requires only one qubit and uses fewer experimental resources as compared to
similar work.

The material of the paper is arranged as follows: Section~\ref{theory}
describes the theoretical background of the variational quantum eigensolver
algorithm, the variational quantum deflation algorithm, the electronic
structure of H$_2$ molecule and the solution of the H$_2$ molecular
Hamiltonian using one
and two qubits. Section~\ref{experiment} describes the  NMR implementation of
the ground and excited state energies  
of the H$_2$ molecule using a two-qubit and a single-qubit
system, respectively.  
Section~\ref{con} contains a few concluding remarks.
\section{Theoretical background}
\label{theory}
\subsection{Quantum variational methods for energy calculations}
\label{sec2a}
Given that quantum systems are described by
Hamiltonians, finding their eigenvalues and eigenvectors is
of paramount importance and becomes increasingly difficult
with increase in system dimensionality.  For a Hamiltonian $H$, an ansatz
state $\ket{\psi(\theta)}$ is chosen with at least one free
parameter $\theta$.  The expectation value of energy
$\epsilon_\theta = \langle
\psi(\theta)|H|\psi(\theta)\rangle $ is minimized over the
parameter $\theta$ to obtain the minimum possible energy.
The lowest expectation value that can be obtained in this way
is the ground state energy of the Hamiltonian itself. In
that case, the state $\ket{\psi(\theta)}$ will be the ground
state of the given Hamiltonian. However, this may not work
as the family of states $\ket{\psi(\theta)}$ may not contain
the ground state! Therefore, a good choice of ansatz is
important.  On a quantum computer, there is an additional
possibility of storing  the quantum state in the quantum
memory and carrying out the entire optimization process as a quantum
computing process.  The above variational
VQE algorithm is a hybrid one, because 
a classical computer is used to update the value of the
parameter $\theta$~\cite{jarrod-njp-2016}.

A variation of the above method can be used to calculate 
the excited
states of the Hamiltonian and the procedure is called
the VQD method~\cite{higgott-quan-19}.
If $\ket{\phi_0}$ is the ground state of the original
Hamiltonian $H$, the VQD Hamiltonian for the first excited
state is given by: 
\begin{equation}
H_1 = H + \beta_0 |\phi_0\rangle \langle \phi_0 |
\end{equation}
where $\beta_0$ is a parameter and is chosen such that the 
variational calculation applied on $H_1$ gives its minimum
energy (such that it is equal to the first excited state of
of the original Hamiltonian $H$). The VQD method can be easily
extended to find the other excited states of $H$ by
appropriately constructing the new Hamiltonian for the
variational purpose.
\subsection{Solving for energies of the H$_2$ molecule}
\label{sec2b}
The Hamiltonian of the H$_2$
molecule in atomic units is given
by~\cite{colless-prx-18,malley-prx-16}:
\begin{equation}
\begin{aligned}
 H = -\sum_i \dfrac{\nabla_{R_i}}{2M_i} -\sum_i
\dfrac{\nabla_{r_i}}{2} +\sum_{i,j>i} \dfrac{Z_i Z_j}{|R_i -
R_j|}\\ -\sum_{i,j>i} \dfrac{Z_i}{|R_i
    - r_j|} +\sum_{i,j>i} \dfrac{1}{|r_i - r_j|}
      \label{eq:h2_hamiltonian}  \end{aligned}
\end{equation}
where $R_i, M_i, Z_i$ denote the position, mass and
charge of the $i^{\text{th}}$ nuclei respectively, and $r_i$
denotes the position of the $i^{\text{th}}$ electron.

The above Hamiltonian is first solved in the Born-Oppenheimer approximation and
then cast into the second quantized form using a specific choice of
$N$-particle basis functions~\cite{malley-prx-16}.  Since quantum
many-body problems are described in the language of second quantization, one
needs a formalism to translate the language of second quantization to a
language that a quantum computer can read. This can be achieved in various
ways, including the Jordan-Wigner transformation, the Bravyi-Kitaev
transformation, and the 
Parity transformation~\cite{malley-prx-16}.  
In this work, we use the Parity
transformation. 
The quantum many-body Hamiltonian can be written in second
quantization as~\cite{malley-prx-16}:
\begin{gather*}
    H = \sum_{\mu,\nu}t_{\mu,\nu}c^{\dagger}_{\mu}c_{\nu} + \sum_{\mu,\nu, \mu' \nu'}V_{\mu,\nu, \mu' \nu'} c^{\dagger}_{\mu} c^{\dagger}_{\nu} c_{\nu'} c_{\mu'} \\
    \intertext{where,}
   t_{\mu,\nu} = \int d\sigma \phi^*_\mu(\sigma)\left( -\dfrac{\nabla_{r}}{2} - \sum_{i} \dfrac{Z_i}{|R_i- r|}\right) \phi_\mu(\sigma)\\ 
     \vspace{1em}\\
     V_{\mu,\nu, \mu' \nu'} = \int d\sigma_1d\sigma_2 \dfrac{\phi^*_\mu(\sigma_1)\phi^*_\nu(\sigma_2)\phi_{\mu'}(\sigma_1)\phi_{\nu'}(\sigma_2)}{|r_1 - r_2|}
 \end{gather*}
where $\sigma_i = (r_i, s_i)$ with $r_i,s_i$ are the position and spin
indices, respectively.  The operators $c_\mu, c_\mu^\dagger$ are the lowering
and raising operators for electron occupation in orbitals. 

To solve the above Hamiltonian on a quantum computer using the VQD algorithm,
all the second quantized operators have to be mapped to the Pauli operators,
and the orbitals have to be mapped to qubits.  Using the Parity transformation,
the many-body H$_2$ Hamiltonian is mapped to the qubit Hamiltonian acting in a
two-qubit state space~\cite{seeley-jcp-12,ganzhorn-pra-19}: 
\begin{equation}
H = a_0 II + a_1 ZI + a_2 IZ + a_3 ZZ + a_4 XX
\label{eq:reduce_hamiltonian} 
\end{equation}
The coefficients $a_i$ encode the information of molecular integrals
pertaining to a specific basis set. 
We have chosen the STO-3G basis in our calculations.
The Qiskit.Chemistry module~\cite{qiskit2024} 
was used to generate the Hamiltonian in the STO-3G
basis, which was then converted to a qubit Hamiltonian 
using the Parity map. The values of the
$a_i$ coefficients corresponding to each Pauli operator
were extracted as a function of different internuclear distances.

In order to use the variational principle for the minimum energy
calculation, we use the Unitary Coupled Cluster Singles and
Doubles (UCCSD) ansatz:
\begin{equation}
\ket{\psi(\theta)} = U(\theta)\ket{\psi_0} \label{eq:anastz} 
\end{equation}
where $\ket{\psi_0}$ is an initial reference state and
$U(\theta)$ is the UCCSD evolution operator. 
The choice of the reference state is the Hartree-Fock (HF)
state, which definitely has a support on the ground state of
the Hamiltonian. After writing the HF state in the Parity
basis, the symmetry in the action of UCCSD operator and the
HF state can be exploited to give a two-qubit UCCSD
operator:
\begin{equation}
U(\theta) = \exp(i\theta XY)
\label{eq:uccsd_2} 
\end{equation}
with the HF state given by $\ket{01}$~\cite{malley-prx-16}.
The action of $U(\theta)$ on $\ket{01}$ can be computed
from:
\begin{equation}
U(\theta)\ket{01} = \exp(i\theta XY)\ket{01}
= \cos{\theta}\ket{01} + \sin{\theta}\ket{10}
\label{eq:uccsd_2_application}  
\end{equation}
The action of
$U(\theta)$ on $\ket{01}$, and then minimization of the
energy expectation leads to the ground state of the
Hamiltonian. The classical minimization process is done
using the Nelder-Mead method given by the minimize function
of the {\em scipy.optimize} module~\cite{olsson-techno-75,scipy}. 
The VQD modified Hamiltonian and the same ansatz
leads to an excited state of the H$_2$ molecule.

Given that we obtained two energy eigenstates of the
Hamiltonian in the space spanned by the vectors
{$\ket{01},\ket{10}$}, 
the other two eigenstates
belong to the space spanned by the other two computational
basis vectors, namely $\vert 00\rangle$ and $\vert 11
\rangle$.
In order to find the other two eigen states,
we take the
initial state to be $\ket{00}$, and apply the ansatz circuit
which performs the evolution $U(\theta)$ (as in the
previous case) given by:
\begin{equation}
U(\theta) = \exp(i\theta XY)\\
U(\theta)\ket{00} = \cos{\theta}\ket{00} - \sin{\theta}\ket{11}
 \label{eq:uccsd_3_application}  
\end{equation}
Since the H$_2$ Hamiltonian of
Eqn.~\eqref{eq:reduce_hamiltonian} is real in the computational
basis, there has to exist a real eigenbasis for the
Hamiltonian with real coefficients of expansion in
the computational basis.  Thus, we do not have to explore the
space where the relative phase between the two computational
basis vectors is complex.  Hence the problem essentially gets
divided into two problems, each of two dimensions.
\subsection{Solving for the  energies
of the H$_2$ molecule using a single qubit}
\label{sec2c}
As is clear from Equation~\ref{eq:uccsd_2_application}  
in the previous section, the UCCSD ansatz only explores a 
two-dimensional
subspace of the entire four-dimensional space. This creates a possibility of
mapping the two-dimensional subspace to a single-qubit space.
If we map the state to involved to a single-qubit state as:
\begin{equation}
\ket{01}\rightarrow \ket{0}, \mbox{and} \quad
\ket{10}\rightarrow \ket{1}
\end{equation}
the UCCSD operator gets
mapped to $\exp(-i\theta Y)$, where $Y$ acts on the 
single-qubit space. 

In order to  map the Hamiltonian, we first evaluate how different
terms of the Hamiltonian act on the basis vectors spanning
the subspace (Table~\ref{first_map}).

\begin{table} \setlength{\tabcolsep}{8pt}
\renewcommand{\arraystretch}{1.5} \caption{ The action of
the individual terms of the Hamiltonian on the states
$\ket{01}$ and $\ket{10}$.} \vspace*{12pt} \centering
\begin{tabular}{rlr|rlr} \hline \hline $II\ket{01}$ & = &
$\ket{01}$  & $II\ket{10}$ & = & $\ket{10}$  \\ $ZI\ket{01}$
& = & $\ket{01}$  & $ZI\ket{10}$ & = & $-\ket{10}$ \\
$IZ\ket{01}$ & = & $-\ket{01}$ & $IZ\ket{10}$ & = &
$\ket{10}$  \\ $ZZ\ket{01}$ & = & $-\ket{01}$ & $ZZ\ket{10}$
& = & $-\ket{10}$ \\ $XX\ket{01}$ & = & $\ket{10}$ &
$XX\ket{10}$ & = & $\ket{01}$  \\ \hline \end{tabular}
\label{first_map} \end{table}

It is evidence from 
Table~\ref{first_map}, that:
\begin{align}
II \longrightarrow I\ \ &\& \ \ ZZ \longrightarrow -I \nonumber \\
ZI \longrightarrow Z\ \ &\& \ \ IZ \longrightarrow -Z \nonumber \\ 
XX &\longrightarrow X
\end{align}
Thus, the diagonalization of the Hamiltonian in the two-dimensional
subspace of the two-qubit space reduces to the diagonalization
of a single-qubit  Hamiltonian given by:
\begin{equation}
H = (a_0 - a_3)I + (a_1 - a_2)Z + a_4X  
\label{eq:1q_reduce_hamiltonian1}
\end{equation}
with the UCCSD ansatz $\exp(-i\theta Y)$ and the initial state $\ket{0}$.

For the remaining two eigenstates, we follow an analogous
procedure to obtain:
\begin{equation}
H = (a_0 + a_3)I + (a_1 + a_2)Z + a_4X    
\label{eq:1q_reduce_hamiltonian2}
\end{equation}
with the UCCSD ansatz $\exp(i \theta Y)$ and the initial state $\ket{0}$.

Diagonalizing these two Hamiltonians will give all
the four excited states of the original two-qubit H$_2$
Hamiltonian. The reduction of the problem to 
a single qubit is
resource efficient from the point of view of its experimental
implementation. Further, since there are only three terms in the
Hamiltonian, the calculation of energy expectation values becomes
simpler. Additionally, the UCCSD ansatz, unlike in the two-qubit
qubit case, is now a simple rotation about the $Y$-axis, which is
easier to implement experimentally 
on any quantum platform. Last but not
the least, the initial state can be directly taken to be the
pseudopure state on an NMR quantum computer,
without any additional requirement to prepare a
reference state. With all these advantages, we have shown
that the entire problem of diagonalization of the two-qubit
H$_2$ molecule can be easily recast as a problem of
diagonalization in a single-qubit space.
\section{Experimental Implementation}
\label{experiment}
\subsection{Calculating energies of the H$_2$
molecule using two NMR qubits} 
\label{sec3a}
\begin{figure}[h]
\centering
\includegraphics[scale=1.0]{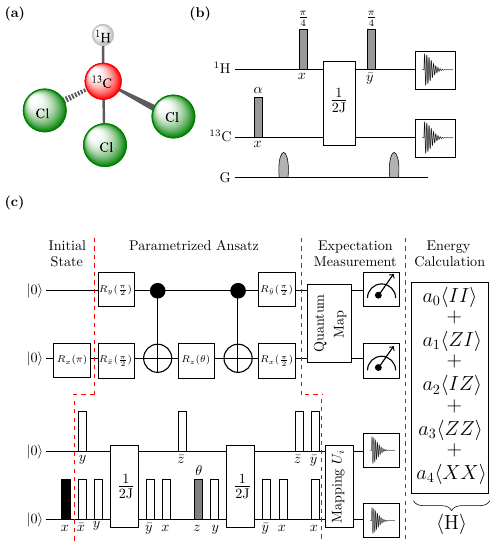}
\caption{(a) Molecular structure of $^{13} \rm C$-labeled
chloroform used as a two-qubit quantum system. (b) NMR pulse
sequence for the PPS $|00\rangle$ state where the value of
the flip angle $\alpha$ is kept fixed at $59.69^{\circ} $,
while $\rm J$ represents the coupling between the  $^{1} \rm
H$ and  $^{13} \rm C$;  the total time evolution is
given by $\frac{1}{2 \rm J}$.
(c) Quantum circuit for the required
state, generated with an optimized value of $\theta$ for
different intermolecular distances. The unfilled rectangles
denote $\pi$ pulses, while the filled rectangle represents a
$\frac{\pi}{2}$ pulse. The flip angles and phases of the
other pulses are written below each pulse, and a  bar over the
phase  denotes negative phase.} \label{male_pps}
\end{figure}

To experimentally
calculate the energies simulated by  VQE
algorithms of the H$_2$ molecule on a four-dimensional quantum
system, the molecule of $^{13} \rm C$ enriched chloroform
dissolved in acetone-D6 was used, with the $^{13} \rm C$ and
$^{1} \rm H$ spins being labeled as qubit $1$ and qubit $2$,
respectively (Fig.~\ref{male_pps}(a)).  NMR experiments
are sensitive only to the deviation density matrix and the
initial state is typically prepared from thermal equilibrium as
a pseudopure state (PPS), which mimics the evolution 
of true pure states~\cite{oliveira-book-07}:
\begin{equation}
\rho_{00} = \frac{1 -\epsilon}{2^3} I_4  + \epsilon |00\rangle \langle 00|
\label{eq:pps deviation}
\end{equation}
where spin polarization $\epsilon \approx 10^{-6}$ and $I_4$ is the $4 \times 4
$ identity operator. We initialized the system in the PPS
$|00\rangle$ using the spatial averaging technique~\cite{oliveira-book-07}, via
a combination of RF pulses and pulsed magnetic gradients. The NMR pulse
sequence for the PPS is given in Fig.~\ref{male_pps}(b).

\begin{table*}
\setlength{\tabcolsep}{16pt} 
\renewcommand{\arraystretch}{1.7}
\caption{Hamiltonian coefficients $a_i, i=0,1,2,3,4$ for
different
internuclear separations (R).}
\vspace*{12pt}
\centering
\begin{tabular}{cccccc}
\hline
\hline
	R ({\AA}) & $a_0$       & $a_1$      & $a_2$       & $a_3$       & $a_4$    \\ \hline 
0.30                              & -0.75374 & 0.80864 & -0.80864 & -0.01328 & 0.16081  \\
0.40                              & -0.86257 & 0.68881 & -0.68881 & -0.01291 & 0.16451  \\ 
0.50                              & -0.94770 & 0.58307 & -0.58307 & -0.01251 & 0.16887  \\ 
0.60                              & -1.00712 & 0.49401 & -0.49401 & -0.01206 & 0.17373  \\ 
0.70                              & -1.04391 & 0.42045 & -0.42045 & -0.01150 & 0.179005 \\ 
0.80                              & -1.06321 & 0.35995 & -0.35995 & -0.01080 & 0.18462  \\ 
0.90                              & -1.07028 & 0.30978 & -0.30978 & -0.00996 & 0.19057  \\ 
1.00                              & -1.06924 & 0.26752 & -0.26752 & -0.00901 & 0.19679  \\ 
1.10                              & -1.06281 & 0.23139 & -0.23139 & -0.00799 & 0.20322  \\ 
1.20                              & -1.05267 & 0.20018 & -0.20018 & -0.00696 & 0.20979  \\ 
1.30                              & -1.03991 & 0.17310 & -0.17310 & -0.00596 & 0.21641  \\ 
1.40                              & -1.02535 & 0.14956 & -0.14956 & -0.00503 & 0.22302  \\ 
1.50                              & -1.00964 & 0.12910 & -0.12910 & -0.00418 & 0.22953  \\ 
1.60                              & -0.99329 & 0.11130 & -0.11130 & -0.00344 & 0.23590  \\ 
1.70                              & -0.97673 & 0.09584 & -0.09584 & -0.00280 & 0.24207  \\ 
1.80                              & -0.96028 & 0.08240 & -0.08240 & -0.00226 & 0.24801  \\ \hline\hline
\end{tabular}
\label{coefficiants}
\end{table*}

\begin{figure}[h]
\includegraphics[scale=1.0]{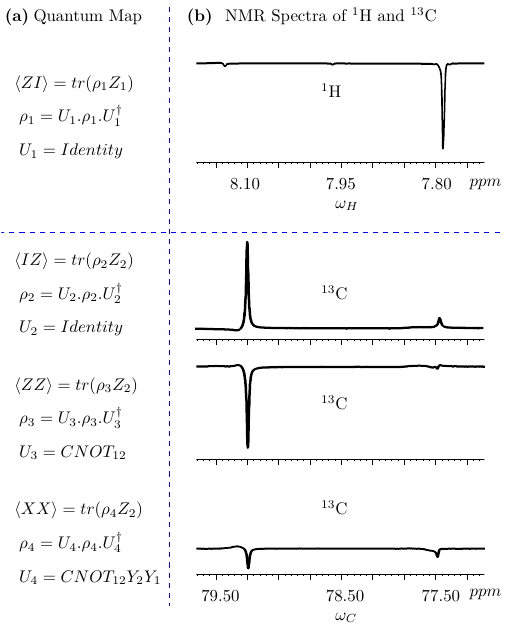}
\caption{(a) Details of the mapping performed for the measurement of the
expectation values $\langle ZI \rangle$, $\langle
IZ\rangle$, $\langle ZZ\rangle$, $\langle XX\rangle$.
The Identity operator $I$ denotes `no operation'.
(b) NMR spectra
of ${}^{1}$H and ${}^{13}$C, displaying the experimentally measured
expectation values of $\langle ZI\rangle$ and 
$\langle IZ\rangle$, $\langle
ZZ\rangle$, $\langle XX\rangle$, respectively, for the ground
state energy calculation, at an intermolecular distance $R=0.70$~{\AA}.}
\label{nmr_expectation} \end{figure}

Experiments were performed on a Bruker Avance III 600-MHz FT-NMR spectrometer
equipped with a TXI probe. Quantum gates, required for the NMR implementation,
were achieved by using specially crafted RF pulses of suitable amplitude,
phase, and duration and nonlocal unitary operations were achieved by free
evolution under the system Hamiltonian. The T$_1$ and T$_2$ relaxation times of
the ${}^{1}$H spin-1/2 nuclei were $\approx 8$ sec and $ \approx3$ sec,
respectively. The T$_1$ and T$_2$ relaxation times of the ${}^{13}$C spin-1/2
nuclei were  $\approx 17$ sec and $\approx 0.5$ sec, respectively.  The
duration of the $\frac{\pi}{2}$ pulses for $^{1} \rm H$  and $^{13} \rm C$
nuclei were $7.14$ $\mu$s at a power level of $19.9$ W, and $12.4$ $\mu$s at a
power level of $237.3$ W, respectively.

In order to experimentally calculate the energies of the
H$_2$ molecule, we need to calculate the expectation values
of $\langle ZI\rangle$, $\langle IZ\rangle$, $\langle
ZZ\rangle$, $\langle XX\rangle$. In an NMR experiment these
expectation values can be calculated by mapping
them onto
the single-qubit Pauli
$Z$ operator. This mapping is easily performed in the
context of an NMR experimental measurement as 
the observed $z$ magnetization of a nuclear spin in
a particular quantum state is proportional to the
expectation value of the Pauli $Z$ operator of the spin in
that state~\cite{gaikwad-pra-18,dileep-pra-19}. For
instance, in order to determine the expectation value
$\langle ZZ\rangle$ for the state 
$\rho=|\psi\rangle\langle \psi|$, we map the state $\rho$ to
$ \rho_1 = U_1 \rho U_1^\dagger$, where $U_1=$CNOT$_{12}$
followed by observing $\langle Z_2\rangle$ for the state
$\rho_1$. The expectation value of $ \langle Z_2 \rangle$
for the state $\rho_1$ is equivalent to observing the
expectation value of $\langle ZZ\rangle$ for the state
$\rho$.

\begin{figure}[h]
\includegraphics[scale=1.0]{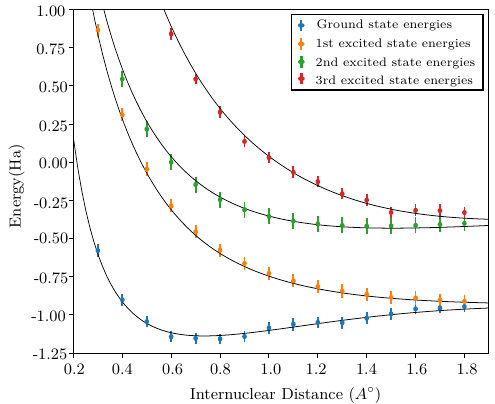}
\caption{Ground and excited state energies of the H$_2$ molecule calculated
using the VQE and VQD algorithms, respectively, over a range of
internuclear separations.  The simulated results are represented by
	joined lines, while the experimentally calculated values (obtained
	using two NMR qubits) are
represented by distinct points with accompanying error bars.  }
\label{h2_energy_spectra} 
\end{figure}

The quantum circuit and corresponding NMR pulse sequence
for the experimental calculation of required expectation
values is given in Fig.~\ref{male_pps}(c). The circuit
comprises three parts (delineated by dashed red lines
in Fig.~\ref{male_pps}(c)): the first part corresponds to 
initializing the
state to the HF state $|01\rangle$, which is achieved by
applying a single-qubit rotation on the $|00\rangle$ state.
The second part of the circuit applies the parameterized ansatz $U(\theta)$ on
the initial state, which is achieved by optimizing the value
of $\theta$  (using the
Nelder-Mead method and the
{\em scipy.optimize} module).  The third part 
of the circuit calculates the
expectation values $\langle ZI\rangle$, $\langle IZ\rangle$,
$\langle ZZ\rangle$, $\langle XX\rangle$ for the
parameterized ansatz, which is achieved by mapping
these expectation values onto
the single-qubit Pauli
$Z$ operator. The experimentally calculated
energies are obtained by adding these expectation values  to
their respective
electronic constants ($a_i, i=0,1,2,3,4$). The numerical values of
the electronic constants are tabulated in 
Table~\ref{coefficiants}, for different values of
the internuclear separation (R) of the H$_2$ molecule.

\begin{figure}[h]
\includegraphics[scale=1.0]{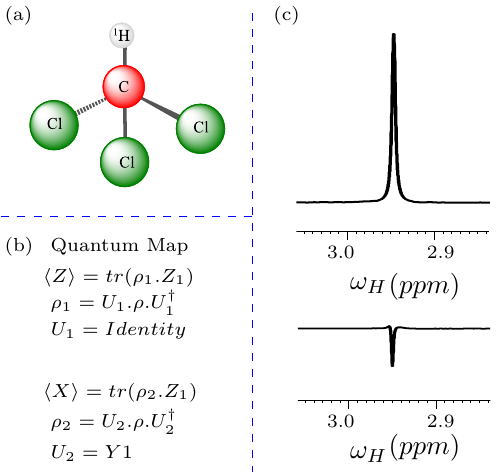}
\caption{(a) Molecular structure of chloroform, used as a
one-qubit quantum system. (b) Mapping details for the
measurement of the expectation values $\langle Z \rangle$ and
$\langle X \rangle$. The identity operator $I$ denotes `no 
operation'. (c)  NMR spectra of ${}^{1}$H, depicting the
experimentally measured expectation values of $\langle
Z\rangle$ and $\langle X \rangle$. } 
\label{male_1qubit}
\end{figure}

The ground state and the
excited state energies of the H$_2$ molecule were
experimentally calculated using VQE and VQD algorithms,
respectively, for sixteen different internuclear distances 
in the range $0.30-1.80$~{\AA}. For each of
these internuclear separations, there are 
five associated electronic
constants $a_0,a_1,a_2,a_3,a_4$ (as given in the H$_2$
molecular Hamiltonian in Eqn.~\eqref{eq:reduce_hamiltonian}).
Fig.~\ref{nmr_expectation}(a) contains the details of
mapping that we have used in the NMR experiment to
calculate  the expectation values required for the
experimentally prepared states.
Fig.~\ref{nmr_expectation}(b) depicts the NMR spectra of
the expectation values $\langle ZI \rangle$, $\langle IZ\rangle$, $\langle
ZZ\rangle$, $\langle XX\rangle$ for the ground state energy
at an internuclear separation of R$=0.70$~{\AA.}.
The ground and excited state energies of the H$_2$ molecule 
have been experimentally calculated using a similar method, for
the thirteen other internuclear separations. The
ground and excited state energies 
have been plotted as a function of 
the internuclear
distances in 
Fig.~\ref{h2_energy_spectra}. As evident from the plot in
Fig.~\ref{h2_energy_spectra}, the simulated and
experimentally measured values agree well, to within
experimental errors. 

\subsection{Calculating energies of the H$_2$
molecule using one NMR qubit}
\label{sec3b}
To experimentally calculate the energies 
of the H$_2$ molecule 
via variational quantum algorithms 
implemented
on a
two-dimensional quantum system, the molecule of chloroform
dissolved in acetone-D6 was used, with  
the ${}^{1}$H spin realizing a single
qubit (Fig.~\ref{male_1qubit}(a)). 
In order to perform the energy calculation, we need
to calculate only two expectation values, namely $\langle Z
\rangle $ and $\langle X \rangle$.
The details of the mapping used for
the experimental calculation are given in
Fig.~\ref{male_1qubit}(b). Fig.~\ref{male_1qubit}(c),
depicts the NMR spectra of the expectation values of
$\langle Z \rangle $, $\langle X \rangle $ for the ground state
energy of the H$_2$ molecule at the internuclear separation
R$=0.70$~{\AA}.

\begin{figure}[h]
\includegraphics[scale=1.0]{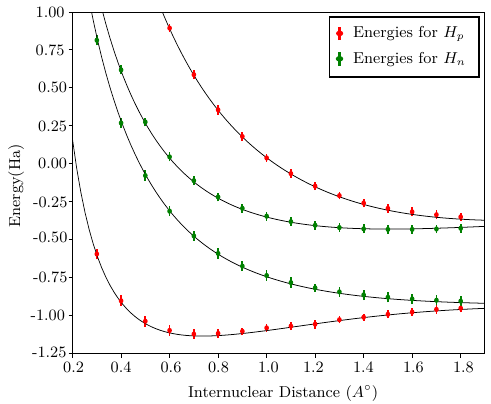}
\caption{Simulations and experimental calculations of the ground and excited
state energies of the H$_2$ molecule, for a range of internuclear
separations.  The simulated results are represented by joined lines,
	while the experimentally calculated values (obtained using
	a single NMR qubit) are represented by distinct
points with accompanying error bars.
}
\label{h2_energy_spectra_1q} 
\end{figure}

The ground and excited state energies for the H$_2$ molecule have been
calculated experimentally for sixteen different internuclear separations. The
ground and excited state energies have been plotted as a function of  the
internuclear separation in Fig.~\ref{h2_energy_spectra_1q}.  The red dashed
curve represents the energy calculation corresponding to the reduced
single-qubit Hamiltonian given in Eqn.~\eqref{eq:1q_reduce_hamiltonian1}. The
green curve represents the energy calculation corresponding to the reduced
single-qubit Hamiltonian given in Eqn.~\eqref{eq:1q_reduce_hamiltonian2}. As is
evident from the plot in Fig.~\ref{h2_energy_spectra_1q}, the simulated and
experimentally measured values agree well to within experimental errors.
Furthermore, it is remarkable that the simulated and experimentally
measured values of the molecular energies in Fig.~\ref{h2_energy_spectra_1q},
are very close to the values obtained in Fig.~\ref{h2_energy_spectra}, 
validating the efficiency of the VQD algorithm using only a single qubit.

The experimental complexity is reduced due to the reduction
of two-qubit system to a one-qubit system. We were hence able to
calculate the energy spectra of  the H$_2$ molecule by measuring
only two expectation values, namely, $\langle Z \rangle $
and $\langle X \rangle$. In the corresponding NMR experiments, 
these expectation values can be
obtained by applying a single RF pulse on the experimentally
prepared state. This is the first experimental demonstration
on an NMR quantum simulator of the calculation of 
energies of the  H$_2$
molecule which requires only a single qubit for its implementation. 

\section{Conclusions}
\label{con}
The ground and excited states of the H$_2$ Hamiltonian were obtained in the
Parity basis using VQE and VQD algorithms, and the results of the
calculations were verified on an NMR quantum simulator.  Further,
the problem of calculation of 
the H$_2$ ground state energy was reduced to a
single-qubit problem which was also verified on the NMR quantum
simulator.  Two NMR qubits were used for the simulation of the
single-qubit problem and the expectation values involved in the
Hamiltonian were obtained  by measuring the single-qubit Pauli $Z$
operator of the experimentally prepared state.  
The experimental implementation of variational
quantum algorithms to calculate the ground and excited state energies
of molecules  is important for the field of quantum chemistry, and our
results are a step forward in this direction.
	
\begin{acknowledgments} All the experiments were performed on a Bruker
Avance-III 600 MHz FT-NMR spectrometer at the NMR Research
Facility of IISER Mohali.  A. acknowledges financial support
from DST/ICPS/QuST/Theme-1/2019/Q-68.  K~.D. acknowledges
financial support from DST/ICPS/QuST/Theme-2/2019/Q-74.
\end{acknowledgments}


%
\end{document}